\documentstyle[12pt,epsf,rotate]{article}
\if@twoside \oddsidemargin 21pt \evensidemargin 59pt \marginparwidth 85pt
\else \oddsidemargin 0pt \evensidemargin 0pt
\fi
\renewcommand{\bibitem}[1]{\\[0.2cm]\vphantom{#1}}
\newcounter{formel1}
\newcounter{formel2}
\newcounter{formel3}
\newcounter{formel4}
\newcounter{frequency}
\newcounter{kappa}

\begin{document}

\title{A Mathematical Model for Attitude Formation by Pair Interactions}

\author{Dirk Helbing \\II. Institute for Theoretical Physics\\
University of Stuttgart, Germany}
\maketitle

\begin{abstract}
\mbox{ }\\[-0.4cm]
Two complementary mathematical models for 
attitude formation are considered: Starting from
the model of {\sc Weidlich} and {\sc Haag} (1983), which assumes
indirect interactions that are mediated by a mean field,
a new model is proposed,
which is characterized by
direct pair interactions. Three types of pair interactions
leading to attitude changes can be found: First, changes by some
kind of avoidance behavior. Second, changes by a readiness for compromises.
Third, changes by persuasion. Different types of behavior are distinguished
by introducing several subpopulations. Representative solutions of the 
model are illustrated by computational results.
\end{abstract}
{\small {\sc Key Words:} attitude formation, 
direct interactions,
indirect interactions, mean field, master equation, probability distribution
of attitudes, persuasion, avoidance behavior, compromises, 
oscillatory attitude changes, 
frequency locking, interacting populations.}\\
{\small {\sc Type of Article:} mathematical model and theory.}\\
{\small {\sc Dimensions and Units:} transition probabilities per unit time
(transition rates); and others.}
                                                                      
\section{Introduction and Summary}

In the field of attitude research there is a broad and substantial
literature available (see e.g. {\sc Fishbein} \& {\sc Ajzen} (1975),
{\sc Petty} \& {\sc Cacioppo} (1986) for an overview). Especially,
some {\em formal} models have been developed like {\sc Osgood} and 
{\sc Tannenbaum}'s (1955) {\em congruity principle},
{\sc Heider}'s (1946) {\em balance theory} or
{\sc Festinger}'s (1957) {\em dissonance theory}. 
These models deal with the {\em stability} of
attitude structures and the probabilities for special types of of attitude
changes. 
\par
In contrast, this paper shall treat general mathematical models for the
change of the {\em fraction} of individuals having a certain attitude $i$.
Such models are of great interest for a {\em quantitative} understanding
of attitude formation, and, especially, for the {\em prognosis} of trends
in the public opinion (e.g. of tendencies in the behavior of voters or
consumers).
{\sc Coleman}'s (1964) {\em diffusion model} was one of the first
quantitative models of this kind. It already distinguished two different
types of attitude changes: First, changes by direct interactions. Second,
externally induced changes. Later,
{\sc Bartholomew} (1967) developed a
{\em stochastic} model for the {\em spread of rumors}. 
\par
The most advanced model has been introduced by {\sc Weidlich} 
({\sc Weidlich} (1972), {\sc Weid\-lich} \& {\sc Haag} (1983)),
which has been seized by some authors in one or another modification
(e.g. {\sc Troitzsch} (1989), {\sc Schweitzer} et. al. (1991)). 
It is based on a {\em master equation} (a stochastic formulation), which
assumes attitude changes to occur with a certain 
{\em probability} per unit time (sect. \ref{MASTER}).
Whereas {\sc Weidlich} and {\sc Haag} (1983) assumed attitude changes by
{\em indirect interactions} (which could only
describe attitude changes via the media
or the spirit of age) (sect. \ref{MASTER}), 
the author thinks that attitude changes by
{\em direct interactions} (due to conversations or discussions)
are of special interest. Therefore, a modified master equation is introduced
in section \ref{Pair}, which allows to examine the effects of 
{\em pair interactions} (dyadic interactions). From this master equation,
certain {\em rate equations} for the temporal development of the
{\em attitude distribution} can be derived (sect.~\ref{Pair}).
\par
Three types of of attitude changes by pair interactions can be found
(sect. \ref{Pair}):
First, an {\em avoidance behavior}, 
causing an individual to change the attitude
if meeting another individual with the 
same attitude (defiant behavior, snob effect).
Second, a readiness for {\em compromises}. Third, the tendency to take over
the opinion of another individual by {\em persuasion}. All three types
of attitude changes may, in the course of time, 
lead to stable equilibrium fractions of the
attitudes, which depend on the {\em preferences} for them.
However, there may also appear more complex attitude changes which could
not be predicted solely by {\em qualitative} 
analyses, for example, {\em oscillatory}
or {\em chaotic} attitude changes. 
\par
In section \ref{Comp}, possible solutions of the model are examined by
considering representative cases. By distinguishing
several {\em subpopulations} (e.g. blue and white collars), different types
of behavior are taken into account. The mutual influence of the
subpopulations is assumed to depend on their mutual sympathy.

\section{Indirect interactions} \label{MASTER}

In the following we shall shortly discuss the modell of {\sc
Weidlich} and {\sc Haag}. For a more detailled description see
{\sc Weidlich} (1972) and {\sc Weidlich} \& {\sc Haag} (1983).
\par
Suppose we have a system consisting of $N$ individuals. These individuals
can normally be divided into $A$ {\em subpopulations} $a$ consisting of
$N_a$ individuals, i.e.
\begin{equation}
 \sum_{a=1}^A N_a = N \, .
\end{equation}
By subpopulations $a$, different social groups (e.g. blue and white collars)
or different characteristic {\em types}
$a$ of behavior are distinguished with respect to the situation of interest
(e.g. the voting behavior).
The $N_a$ individuals of each subpopulation $a$ are
distributed over several {\em states}
\begin{equation}
 \vec{x} \in \{\vec{x}_1,\dots,\vec{x}_{S_a}\} \, .
\end{equation}
Being in state
\begin{equation}
 \vec{x} := (\vec{r},i)
\end{equation}
means being at place $\vec{r}$ and having the attitude $i$ concerning
the question of interest (e.g. ``Which political party $i$ would you
vote for, if there were elections today?''). 
If the {\em occupation number} $n_{\vec{x}}^a$
denotes the number of individuals of type $a$ being in
state $\vec{x}$, we have the relation
\begin{equation}
 \sum_{\vec{x}} n_{\vec{x}}^a \equiv \sum_{s=1}^{S_a}
 n_{\vec{x}_s}^a = N_a \, .
\end{equation}
Let
\begin{equation}
 \vec{n} := (n_{\vec{x}_1}^1,\dots,
n_{\vec{x}_{S_1}}^1;\dots;n_{\vec{x}_1}^A,\dots,n_{\vec{x}_{S_A}}^A) 
 = (\vec{n}^1;\dots;\vec{n}^A)
\end{equation}
be the vector consisting of all occupation numbers $n_{\vec{x}}^a$.
This vector is called the {\em socio-configuration}, since it contains
all information about the distribution of the $N$ individuals over
the states $\vec{x}_s$. 
$P(\vec{n},t)$ shall denote the {\em probability} to find the 
socio-configuration $\vec{n}$ at time $t$. This implies
\begin{equation}
 0 \le P(\vec{n},t) \le 1 \qquad \mbox{and} \qquad
\sum_{\vec{n}} P(\vec{n},t) = 1 \, .
\end{equation}
If transitions
from socio-configuration $\vec{n}$ to $\vec{n}'$
occur with a probability of $P(\vec{n}',t+\Delta t|\vec{n},t)$ 
during a short time interval
$\Delta t$, we have a {\em (relative) transition rate} of
\begin{equation}
 w(\vec{n}',\vec{n};t) 
= \lim_{\Delta t \rightarrow 0}
\frac{P(\vec{n}',t+\Delta t|\vec{n},t)}{\Delta t} \, .
\end{equation}
The {\em absolute} transition rate of changes
from $\vec{n}$ to $\vec{n}'$ is the product 
$w(\vec{n}',\vec{n};t)P(\vec{n},t)$ of the probability
$P(\vec{n},t)$ to have the configuration $\vec{n}$
and the {\em relative} transition rate $w(\vec{n}',\vec{n};t)$
if having the configuration $\vec{n}$. Whereas the {\em inflow} into
$\vec{n}$ is given as
the sum over all absolute transition rates of changes from an {\em arbitrary}
configuration $\vec{n}'$ to $\vec{n}$, the {\em outflow} from $\vec{n}$
is given as the sum over all absolute transition rates of changes
from $\vec{n}$ to {\em another} configuration $\vec{n}'$. Since the
temporal change of the probability $P(\vec{n},t)$ is determined
by the inflow into $\vec{n}$ reduced by the outflow from $\vec{n}$, we
find the {\em master equation} 
\begin{eqnarray}
 \frac{d}{dt} P(\vec{n},t) &=&
\mbox{inflow into $\vec{n}$ } \qquad \quad
- \mbox{ outflow from $\vec{n}$} \nonumber \\
&=& \sum_{\vec{n}'} w(\vec{n},\vec{n}';t)P(\vec{n}',t)
- \sum_{\vec{n}'} w(\vec{n}',\vec{n};t)P(\vec{n},t)
\label{master}
\end{eqnarray}
({\sc Haken} (1983)).
\par
{\sc Weidlich} and {\sc Haag} (1983)
have assumed the individuals to change 
from state $\vec{x}$ to state $\vec{x}'$
with a transition rate of
$w_a(\vec{x}',\vec{x};\vec{n};t)$
{\em independently} of each other. Such changes
correspond to transitions of the socio-configuration from $\vec{n}$
to
\begin{equation}
 \vec{n}_{\vec{y}}^a{}_{\vec{x}}^a 
:= (n_{\vec{x}_1}^1,\dots,n_{\vec{x}_{S_1}}^1;\dots;
n_{\vec{x}_1}^a,\dots,
(n_{\vec{y}}^a +1),\dots,(n_{\vec{x}}^a - 1),\dots,
n_{\vec{x}_{S_a}}^a;\dots;n_{\vec{x}_1}^A,\dots,n_{\vec{x}_{S_A}}^A)
\end{equation}
with a transition rate of
$n_{\vec{x}}^a w_a(\vec{x}',\vec{x};\vec{n};t)$, which
is proportional to the number $n_{\vec{x}}^a$ of individuals who can leave
the state $\vec{x}$.
For $w(\vec{n}',\vec{n};t)$ we therefore have
the relation
\begin{equation}
 w(\vec{n}',\vec{n};t) = \left\{
\begin{array}{ll}
n_{\vec{x}}^a w_a(\vec{x}',\vec{x};\vec{n};t)&\mbox{if } 
\vec{n}' = \vec{n}_{\vec{y}}^a{}_{\vec{x}}^a \\
0 & \mbox{else\,.}
\end{array} \right.
\label{w}
\end{equation}
Assumption (\ref{w}) has 
sometimes been doubted to be suitable for describing social processes, since
it handles the interactions of individuals in an {\em indirect} way
via the dependence on the occupation numbers $n_{\vec{x}}^a$.
That means, the individual interactions are mediated by the {\em mean field}
of the socio-configuration, which could model 
attitude changes via the media or the spirit of the age. Figure 1 illustrates
the $2N$ indirect interactions of $N$ individuals.
\par
In contrast, this article will deal with {\em direct} interactions of 
individuals, which are a realistic model for attitude changes in
conversations or discussions. 
We shall consider {\em pair interactions} only, because
they are the most important ones. However, it is also possible to develop
models for simultaneous interactions of an arbitrary number of individuals
({\sc Helbing} (1991a)). Figure 2 depicts the $N(N-1)$ direct
pair interactions of $N$ individuals.

\section{Direct pair interactions} \label{Pair}

A model describing direct pair
interactions is given by the master equation (\ref{master}),
when taking the special transition rates 
\begin{equation}
 w(\vec{n}',\vec{n};t) = \left\{
\begin{array}{ll}
n_{\vec{x}}^a n_{\vec{y}}^b \widetilde{w}_{ab}(\vec{x}',\vec{y}';
\vec{x},\vec{y};t) & \mbox{if } \vec{n}' 
=\vec{n}_{\vec{x}'}^{a}{}_{\vec{y}'}^{b}{}_{\vec{x}}^a{}_{\vec{y}}^b \\
0 & \mbox{else\,,}
\end{array} \right.
\label{Raten}
\end{equation}
where
\begin{equation}
 \vec{n}_{\vec{x}'}^{a}{}_{\vec{y}'}^{b}{}_{\vec{x}}^a{}_{\vec{y}}^b
:= (\dots,(n_{\vec{x}'}^{a}+1),\dots,
(n_{\vec{x}}^a -1),\dots,(n_{\vec{y}'}^{b}+1),
\dots,(n_{\vec{y}}^b -1),\dots) \, .
\end{equation}
$\widetilde{w}_{ab}(\vec{x}',\vec{y}';\vec{x},\vec{y};t)$ 
is the (relative) transition rate for
two individuals of types $a$ and $b$ to change their states from
$\vec{x}$ and $\vec{y}$ to $\vec{x}'$ and $\vec{y}'$ due to direct
interactions. The corresponding transition rate
$n_{\vec{x}}^a n_{\vec{y}}^b 
\widetilde{w}_{ab}(\vec{x}',\vec{y}';\vec{x},\vec{y};t)$ for changes of the
configuration from $\vec{n}$ to
$\vec{n}_{\vec{x}'}^{a}{}_{\vec{y}'}^{b}{}_{\vec{x}}^a{}_{\vec{y}}^b$ is
again proportional to the numbers $n_{\vec{x}}^a$ 
and $n_{\vec{y}}^b$ of individuals which can leave
the states $\vec{x}$ and $\vec{y}$. 
\par
Usually, one is mostly interested in the expected {\em fractions}
\begin{equation}
 P_a(\vec{x},t) := \frac{\langle n_{\vec{x}}^a \rangle}{N_a}
\end{equation}
of individuals of type $a$ who are in state $\vec{x}$.
Here, $\langle n_{\vec{x}}^a \rangle$
denotes the mean value
\begin{equation}
 \langle n_{\vec{x}}^a \rangle := \sum_{\vec{n}} n_{\vec{x}}^a
P(\vec{n},t) 
\end{equation}
of the occupation number $n_{\vec{x}}^a$. For $P_a(\vec{x},t)$ we have the
relations 
\begin{equation}
0\le P_a(\vec{x},t)\le 1 \qquad \mbox{and} \qquad \sum_{\vec{x}}
P_a(\vec{x},t) = 1 \, .
\label{nor}
\end{equation}
$P_a(\vec{x},t)$ can be interpreted as the probability distribution of state
$\vec{x}$ within subpopulation $a$. Since $N_a/N$ is the fraction of
individuals of type $a$, the probability distribution $P(\vec{x},t)$
of state $\vec{x}$ in the {\em entire} population is given by
\begin{equation}
 P(\vec{x},t) = \sum_a \frac{N_a}{N} P_a(\vec{x},t) \, .
\end{equation}
The temporal development of $P_a(\vec{x},t)$ obeys the 
{\sc Boltzmann} equation ({\sc Boltzmann} (1964))\\
\addtocounter{equation}{1}
\setcounter{formel1}{\theequation}
\parbox{14.5cm}{\begin{eqnarray*}
\frac{d}{dt}P_a(\vec{x},t) &=& 
\sum_{b}\sum_{\vec{x}'}\sum_{\vec{y}} \sum_{\vec{y}'}  
w_{ab}(\vec{x},\vec{y}';\vec{x}',\vec{y};t)P_a(\vec{x}',t)P_b(\vec{y},t) 
 \nonumber \\
&-&  \sum_{b}\sum_{\vec{x}'}\sum_{\vec{y}} \sum_{\vec{y}'}  
w_{ab}(\vec{x}',\vec{y}';\vec{x},\vec{y};t)P_a(\vec{x},t)P_b(\vec{y},t) 
\end{eqnarray*}}\hfill
\parbox{1.2cm}{\begin{displaymath}
\begin{array}{r}
\rule[-4mm]{0cm}{1.3cm}(\theequation{}\mbox{a})\\
\rule[-4mm]{0cm}{1.3cm}(\theequation{}\mbox{b})\\
\end{array}
\end{displaymath}}
with
\begin{equation}
 w_{ab} := N_b \widetilde{w}_{ab} \, ,
\end{equation}
as long as the absolute values of the (co)variances 
\begin{equation}
 \sigma_{\vec{x}}^a{}_{\vec{x}'}^b 
:= \sum_{\vec{n}} (n_{\vec{x}}^a - \langle n_{\vec{x}}^a \rangle)
(n_{\vec{x}'}^b - \langle n_{\vec{x}'}^b \rangle) P(\vec{n},t)
= \langle n_{\vec{x}}^a n_{\vec{x}'}^b \rangle 
 - \langle n_{\vec{x}}^a \rangle\langle n_{\vec{x}'}^b \rangle 
\end{equation}
are small compared to $\langle n_{\vec{x}}^a \rangle \langle n_{\vec{x}'}^b
\rangle$, and if $(N_a-1)/N_a \approx 1$ (see {\sc Helbing} (1991a)). 
These conditions are usually fulfilled
in case of a great population size $N$. Equation (\arabic{formel1})
can be also understood as {\em exact} equation for the {\em most
probable} individual behavior.
\par
The interpretation of (\arabic{formel1}) is similar to the one of the
master equation (\ref{master}). Again, the temporal change
of the fraction $P_a(\vec{x},t)$ of individuals being in state $\vec{x}$
is given by the inflow (\arabic{formel1}a) into state $\vec{x}$ reduced by
the outflow (\arabic{formel1}b) from state $\vec{x}$:
\begin{equation}
 \frac{d}{dt} P_a(\vec{x},t) = \mbox{inflow into $\vec{x}$ --
outflow from $\vec{x}$\,.}
\end{equation}
The inflow [resp. outflow] is given by all transitions, where individuals
of type $a$ change their states from arbitrary states $\vec{x}'$ to
state $\vec{x}$ [resp. from state $\vec{x}$ to arbitrary states $\vec{x}'$]
due to pair interactions with other individuals of arbitrary type $b$,
who change their arbitrary states $\vec{y}$ to arbitrary states $\vec{y}'$.
These pair interactions occur with a frequency proportional to the
fractions $P_a$ and $P_b$ of the states which are subject to a change.
Since $b$, $\vec{x}'$, $\vec{y}$ and $\vec{y}'$ are arbitrary, one has to 
carry out a summation over them.
\par
In the following we shall consider the cases of 
{\em spatially homogeneous} or {\em local}
attitude formation, which are independent of $\vec{r}$.
Equation (\arabic{formel1})
has then the form
\begin{equation}
 \frac{d}{dt}P_a(i,t) = \sum_b \sum_{i'}\sum_j
\bigg[ w_{ab}^*(i;i',j;t)P_a(i',t)P_b(j,t)
- w_{ab}^*(i';i,j;t)P_a(i,t)P_b(j,t) \bigg] \, ,
\label{mein}
\end{equation}
where the abbreviation
\begin{equation}
 w_{ab}^*(.;.,.;t) :=  \sum_{j'} w_{ab}(.,j';.,.;t)
\end{equation}
has been used. $w_{ab}^*(i';i,j;t)$ is the rate for individuals of type
$a$ to change from attitude $i$ to attitude $i'$ due to interactions with 
individuals of type $b$ having the attitude $j$.
For the description or prediction of concrete situations
the rates $w_{ab}^*$ have to be determined empirically
(see {\sc Helbing} (1991b)). In the following,
we shall instead examine the possible solutions of (\ref{mein}) by considering
representative examples.
Only three kinds of interaction contribute to the
temporal change of $P_a(i,t)$:\\
\addtocounter{equation}{1}
\setcounter{formel2}{\theequation}
\parbox{14.5cm}{\begin{eqnarray*}
i' &\stackrel{i}{\longleftarrow} & i \quad (i'\ne i) \\
i' &\stackrel{j}{\longleftarrow} & i \quad (j\ne i,\, i'\ne i,\,i' \ne j) \\
j &\stackrel{j}{\longleftarrow} & i  \quad (j\ne i)
\end{eqnarray*}}\hfill
\parbox{1.2cm}{\begin{displaymath}
\begin{array}{r}
\rule[-4mm]{0cm}{0.8cm}(\theequation.1)\\
\rule[-4mm]{0cm}{0.8cm}(\theequation.2)\\
\rule[-4mm]{0cm}{0.8cm}(\theequation.3)
\end{array}
\end{displaymath}}
The interpretation is obviously the following:
\begin{enumerate}
\item[1.] The first term describes some 
kind of {\em avoidance behavior,} causing
an individual to change the opinion from $i$ to $i'$
with a certain probability, when meeting
another individual originally having the same opinion $i$.
Such defiant behavior is e.g. known as snob effect.
\item[2.] The second term represents the readiness to change the opinion
from $i$ to $i'\ne j$ (in the following called a {\em compromise}), 
if meeting another individual
having a different opinion $j\ne i$. This behavior will be found, if attitude
$i$ cannot be maintained when confronted with attitude $j$, but attitude
$j$ is not satisfying (for this person) either.
\item[3.] The third term describes the tendency to take over the opinion $j$
of another individual by {\em persuasion.}
\end{enumerate}
According to this classification, the transition rates $w_{ab}^*(i';i,j)$
can be splitted into three contributions:\\
\addtocounter{equation}{1}
\setcounter{formel3}{\theequation}
\parbox{14.5cm}{\begin{eqnarray*}
 w_{ab}^*(i';i,j;t) &=& w_{ab}^{*1}(i';i;t) \delta_{ij} \\
 &+& w_{ab}^{*2}(i';i,j;t) \\
 &+& w_{ab}^{*3}(j;i;t)\delta_{ji'} \, ,
\end{eqnarray*}}\hfill
\parbox{1.2cm}{\begin{displaymath}
\begin{array}{r}
\rule[-4mm]{0cm}{0.8cm}(\theequation.1)\\
\rule[-4mm]{0cm}{0.8cm}(\theequation.2)\\
\rule[-4mm]{0cm}{0.8cm}(\theequation.3)
\end{array}
\end{displaymath}}
where 
\begin{equation}
 \delta_{ij} := \left\{
\begin{array}{ll}
1 & \mbox{if } i=j \\
0 & \mbox{if } i \ne j
\end{array}\right.
\end{equation}
is the {\sc Kronecker} function and
\begin{equation}
 w_{ab}^{*1}(i,i)=0,\quad w_{ab}^{*2}(i;i,j)=0, 
\quad w_{ab}^{*2}(j;i,j)=0, \quad
w_{ab}^{*2}(i';i,i)=0, \quad w_{ab}^{*3}(i;i)=0 \, .
\label{null}
\end{equation}
Equation (\ref{mein}) has now the form\\
\addtocounter{equation}{1}
\setcounter{formel4}{\theequation}
\parbox{14.5cm}{\begin{eqnarray*}         
\frac{d}{dt}P_a(i,t) &=& 
\sum_{b}\sum_{i'} 
\bigg[w_{ab}^{*1}(i;i';t)P_a(i',t)P_b(i',t) 
- w_{ab}^{*1}(i';i;t)P_a(i,t)P_b(i,t)\bigg] \\
&+& \sum_{b}\sum_{i'}\sum_{j} 
\bigg[w_{ab}^{*2}(i;i',j;t)P_a(i',t)P_b(j,t) 
- w_{ab}^{*2}(i';i,j;t)P_a(i,t)P_b(j,t)\bigg] \\
&+& \sum_{b}\sum_{j} 
\bigg[w_{ab}^{*3}(i;j;t)P_a(j,t)P_b(i,t) 
- w_{ab}^{*3}(j;i;t)P_a(i,t)P_b(j,t)\bigg] \, . 
\end{eqnarray*}}                          
\hfill
\parbox{1.2cm}{\begin{displaymath}
\begin{array}{r}
\rule[-4mm]{0cm}{1.2cm}(\theequation.1)\\
\rule[-4mm]{0cm}{1.2cm}(\theequation.2)\\
\rule[-4mm]{0cm}{1.2cm}(\theequation.3)
\end{array}
\end{displaymath}}
$w_{ab}^{*k}$ ($k=1,2,3$) are the contributions to the transition
rates $w_{ab}^*$ by interactions of type~$k$.

\section{Computer solutions} \label{Comp}

In order to obtain concrete results, we have to make some 
plausible specifications
of the model. Let $\kappa_{ab}$ be the degree of {\em sympathy} which 
individuals of type $a$ feel towards individuals of type $b$. Then one expects
the following: Whereas the 
rate $w_{ab}^{*2}$ of the readiness for compromises and the rate
$w_{ab}^{*3}$ of the tendency to take over another opinion will be increasing
with $\kappa_{ab}$,
the rate $w_{ab}^{*1}$ of the avoidance behavior will be decreasing
with $\kappa_{ab}$.
This functional dependence can e.g. be described by\\
\addtocounter{equation}{1}
\setcounter{frequency}{\theequation}
\parbox{14.5cm}{\begin{eqnarray*}
 w_{ab}^{*1}(i';i;t) &:=& \nu_{ab}^1(t) R_a(i',i;t)(1 - \delta_{ii'}) \\
 w_{ab}^{*2}(i';i,j;t) &:=& \nu_{ab}^2(t) R_a(i',i;t)(1-\delta_{ii'})
 (1-\delta_{ij})(1-\delta_{ji'})  \\
 w_{ab}^{*3}(j;i;t) &:=& \nu_{ab}^3(t) R_a(j,i;t)(1-\delta_{ij}) 
\end{eqnarray*}}\hfill
\parbox{1.2cm}{\begin{displaymath}
\begin{array}{r}
\rule[-4mm]{0cm}{0.8cm}(\theequation.1)\\
\rule[-4mm]{0cm}{0.8cm}(\theequation.2)\\
\rule[-4mm]{0cm}{0.8cm}(\theequation.3)
\end{array}
\end{displaymath}}
with\\
\addtocounter{equation}{1}
\setcounter{kappa}{\theequation}
\parbox{14.5cm}{\begin{eqnarray*}
 \nu_{ab}^1(t) &:=& \nu_a^1 (1-\kappa_{ab}) \\
 \nu_{ab}^2(t) &:=& \nu_a^2 \kappa_{ab} \\
 \nu_{ab}^3(t) &:=& \nu_a^3 \kappa_{ab} 
\end{eqnarray*}}\hfill
\parbox{1.2cm}{\begin{displaymath}
\begin{array}{r}
\rule[-4mm]{0cm}{0.8cm}(\theequation.1)\\
\rule[-4mm]{0cm}{0.8cm}(\theequation.2)\\
\rule[-4mm]{0cm}{0.8cm}(\theequation.3)
\end{array}
\end{displaymath}}
$(0 \le \kappa_{ab} \le 1)$. The factors
$(1-\delta_{ij})$ are due to (\ref{null}). $\nu_a^k$ is a measure for
the rate of attitude changes of type $k$ within subpopulation $a$, i.e.
a {\em flexibility parameter}. 
The rates $\nu_{ab}^k(t)$ can be understood as product of the
{\em contact rate} $\nu_{ab}(t)$ of an individual of subpopulation~$a$ 
with individuals of subpopulation~$b$,
and the frequency $f_{ab}^k(t)$ of having an interaction of type $k$ 
in case of a contact:
\begin{equation}
 \nu_{ab}^k(t) = \nu_{ab}(t) f_{ab}^k(t) \, .
\end{equation}
$\nu_{ab}^k$ has the effect of a {\em coupling coefficient}
and determines, how strong attitude
changes of kind $k$ within subpopulation $a$ depend on the attitude
distribution $P_b(j,t)$ within subpopulation $b$. 
Especially $\nu_{ab}^k=0$ for $k=1,2,3$ implies, that
the dynamic development of $P_a(i,t)$ is completely independent of
$P_b(j,t)$ (see fig. 3).
\par 
For the case of two subpopulations we explicitly have the sympathy matrix
\begin{equation}
 \underline{\kappa} \equiv \bigg( \kappa_{ab} \bigg) = \left(
\begin{array}{cc}
1 & \kappa_{12} \\
\kappa_{21} & 1
\end{array}\right) \, ,
\end{equation}
if the sympathy between individuals belonging to the same subpopulation
is assumed to be maximal ($\kappa_{11}=1=\kappa_{22}$). Then,
the abbreviation
\begin{equation}
 \underline{\kappa}^c_d := \left(
\begin{array}{cc}
1 & c \\
d & 1
\end{array} \right)
\end{equation}
can be used. Figures 3 to 5 ($\nu_1^3=(S_1-1) = 3$, $\nu_2^3 = (S_2 - 1) = 2$,
$\nu_a^1=\nu_a^2=0$, $\underline{R}_1
=\underline{R}^4$, $\underline{R}_2 = \underline{R}^3$) 
demonstrate the effects of variations of
$\underline{\kappa}^c_d$. Attitudes of subpopulation $a=1$ are represented by 
solid lines, attitudes of subpolulation $a=2$ by broken lines. Figure~3 
($\underline{\kappa}=\underline{\kappa}^0_0$) is simulated {\em without} any 
coupling of the subpopulations, figure 4 
($\underline{\kappa}=\underline{\kappa}^0_1$)
with {\em asymmetric} coupling and figure~5 
($\underline{\kappa}=\underline{\kappa}^1_1$) with {\em mutual} coupling.
\par
$R_a(i',i;t)$ is a measure for the {\em readiness
to change the attitude} from $i$ to $i'$ 
for an individual of type $a$, if an attitude change takes place.
We shall assume 
\begin{equation}
 R_a(i,i';t) = \frac{1}{S_a-1} - R_a(i',i;t) \quad 
\mbox{with} \quad 0\le R_a(i',i;t)\le \frac{1}{S_a -1} \, ,
 \quad \mbox{if } i \ne i' 
\label{inverse}
\end{equation}
($S_a = \mbox{number}$ of different attitudes within subpopulation $a$). 
That means, the readiness to change from opinion $i'$ to $i$
will be the greater, the lower the readiness for the inverse change
from $i$ to $i'$ is. 
\begin{equation}
 P_{a,i}(t) :=  \sum_{i'(\ne i)} R_a(i,i';t) 
\end{equation}
can be interpreted as degree of {\em preference} for opinion $i$
(see figures 6 and 7). 
Since $R_a(i,i;t)$ is arbitrary (see (\arabic{frequency}),
(\ref{inverse})), we may define 
\begin{equation}
 R_a(i,i;t) := P_{a,i}(t) \, .
\label{PRO}
\end{equation}
\par
Usually, the equilibrium fraction will 
be growing with the
preference $P_{a,i}$: see figure 6 ($\nu_a^k=S_a - 1 =2$,
$\underline{\kappa}=\underline{\kappa}^0_1$, $\underline{R}_1
=\underline{R}^1$, $\underline{R}_2=\underline{R}^2$), where
a solid line represents the highest preference, a dotted
line the lowest preference, and a broken line medium preference. 
For the special case
\begin{equation}
 P_{a,i} = (S_a -1) \frac{1}{2(S_a -1)} = \frac{1}{2} \, ,
\end{equation}
one of the stationary solutions is
\begin{equation}
 P_a(i) = \frac{1}{S_a} \, .
\end{equation}
This is because of
\begin{equation}
 \sum_{i'(\ne i)} R_a(i',i) = \sum_{i'(\ne i)} 
\left( \frac{1}{S_a -1} - R_a(i,i') \right)
 = 1 - P_{a,i} = \frac{1}{2} = P_{a,i} = \sum_{i'(\ne i)} R_a(i,i')
\end{equation}
and equations (\arabic{formel4}) to (\arabic{kappa}).
An illustration of this case is
given in figure 7 ($\nu_a^k=S_a - 1 =2$,
$\underline{\kappa}=\underline{\kappa}^0_1$, $\underline{R}_a
=\underline{R}^3$).
\par
In the following, we shall often consider the 
situation of three different opinions ($S_a=3$):
\begin{equation}
 \underline{R}_a \equiv \bigg( R_a(i',i) \bigg) := \left(
\begin{array}{ccc}
P_{a,1} & \mbox{$\frac{1}{S_a-1}$} - A_a & \mbox{$\frac{1}{S_a-1}$}-B_a \\
A_a & P_{a,2} & \mbox{$\frac{1}{S_a-1}$}-C_a \\
B_a & C_a & P_{a,3} 
\end{array}\right) \, .
\end{equation}
($A_a$ is the readiness to change from attitude 1 to attitude 2,
$B_a$ to change from 1 to 3, and $C_a$ to change from 2 to 3.)
Especially, for
\begin{equation}
 \underline{R}_a = \underline{R}^1 := \left(
\begin{array}{ccc}
0.7 & 0.35 & 0.35 \\
0.15 & 0.5 & 0.35 \\
0.15 & 0.15 & 0.3 
\end{array}\right) \, , 
\end{equation}
attitude 1 has the highest preference ($P_{a,1}=0.7$), attitude 3 has
the lowest preference ($P_{a,3}=0.3$), and attitude 2 has medium preference
($P_{a,2}=0.5$). For
\begin{equation}
 \underline{R}_a = \underline{R}^2 := \left(
\begin{array}{ccc}
0.5 & 0.15 & 0.35 \\
0.35 & 0.7 & 0.35 \\
0.15 & 0.15 & 0.3 
\end{array}\right) \, , 
\end{equation}
attitude 3 has again the lowest preference ($P_{a,3}=0.3$), 
but attitude 2 has the highest preference
($P_{a,2}=0.7$), and attitude 1 has medium preference ($P_{a,1}=0.5$).
If
\begin{equation}
 \underline{R}_a = \underline{R}^3 := \left(
\begin{array}{ccc}
1/2   & 0   & 1/2   \\
1/2   & 1/2   & 0   \\
0   & 1/2   & 1/2 
\end{array}\right) \, , 
\end{equation}
all attitudes have the same preference ($P_{a,i}=1/2$). An analogous
case for 4 attitudes is given by
\begin{equation}
 \underline{R}_a = \underline{R}^4 := \left(
\begin{array}{cccc}
1/2 & 0   & 1/6 & 1/3   \\
1/3   & 1/2 & 0   & 1/6 \\
1/6 & 1/3   & 1/2 & 0   \\
0   & 1/6 & 1/3   & 1/2
\end{array}\right) \, .
\end{equation}
However, the case of three different opinions represents the prototype of 
opinion formation, since the attitudes concerning a special question can
be classified within the schemes ``positive'', ``negative'', ``neutral''
or ``pro'', ``contra'', ``compromise''. 

\subsection{Avoidance behavior}

In this section we shall examine the effect of avoidance behavior
alone (i.e. $\nu_a^1 = S_a -1 =2$, $\nu_a^2=\nu_a^3=0$). Figures 8 and
9 compare the case, where both subpopulations prefer attitude 1
(Fig. 8: $\underline{\kappa}=\underline{\kappa}^{0.5}_{0.5}$,
$\underline{R}_a=\underline{R}^1$) and where they prefer different attitudes
(Fig. 9: $\underline{\kappa}=\underline{\kappa}^{0.5}_{0.5}$,
$\underline{R}_1=\underline{R}^1$, $\underline{R}_2=\underline{R}^2$).
The attitude $i$ with the highest preference is 
represented by a solid line, the attitude with the
lowest preference by a dotted line, and the one with
medium preference by a broken line. Obviously, 
the fraction of the most prefered attitude is decreasing,
if both subpopulations favour the same
attitude (see fig.~8). If they prefer different attitudes,
it is growing (see fig. 9), since there are less situations of 
avoidance then.  
\par
Avoidance behavior is e.g. known as snob effect. It also
occurs in the case of hostile groups.

\subsection{Readiness for compromises}

Let us now consider the effect of persuasion in combination with
a readiness for compromises
(see figures 10 and 11: $\nu_a^1=0$, $\nu_a^2=\nu_a^3=S_a -1 =2$). 
For the effect of 
persuasion {\em without} readiness for compromises 
(i.e. $\nu_a^1 = \nu_a^2 = 0$, $\nu_a^3 = S_a - 1 = 2$)
see the corresponding figures 12 and 13.
Again, we shall compare the case, 
where both subpopulations prefer attitude 1
(Fig. 10: $\underline{\kappa}=\underline{\kappa}^{1}_{1}$,
$\underline{R}_a=\underline{R}^1$), with the case, 
where they prefer different attitudes
(Fig.~11: $\underline{\kappa}=\underline{\kappa}^{1}_{1}$,
$\underline{R}_1=\underline{R}^1$, $\underline{R}_2=\underline{R}^2$).
If both subpopulations prefer the same attitude, this attitude will be
the only surviving one (see fig. 10). The readiness for compromises will
have little influence then (compare to fig. 12:
$\underline{\kappa}=\underline{\kappa}^{1}_{1}$,
$\underline{R}_a=\underline{R}^1$). However, if the subpopulations
favour different attitudes, the compromise (the 3rd attitude, represented
by the dotted line) will be chosen by a certain fraction of individuals
(see fig. 11). According to this, the competition between several attitudes
will lead to a greater variety of attitudes, 
if there is a readiness for compromises. Without this readiness ($\nu_a^2=0$),
the competing attitudes will be the only surviving ones (see fig. 13:
$\underline{\kappa}=\underline{\kappa}^{1}_{1}$,
$\underline{R}_1=\underline{R}^1$, $\underline{R}_2=\underline{R}^2$).
However, they will survive in {\em both} subpopulations (at least, if
$w_{ab}^{*3}$ is time independent).

\subsection{Effects of persuasion}

This section will deal with the tendency to take over the opinion 
$j$ of another individual (i.e. $\nu_a^3=S_a -1$, $\nu_a^1=\nu_a^2=0$). Typical
cases are depicted in figures 12 and 13. However, 
there are also {\em oscillations} possible as shown in figures 3, 4 and 14
($\underline{\kappa}=\underline{\kappa}^1_1$, $\underline{R}_a
= \underline{R}^3$). In order to understand this behavior, let us consider
the case of a single subpopulation $(A=1)$ or of
uncoupled subpopulations ($\kappa_{ab}=0$ for all $a, b$ with
$a\ne b$). From equations (\arabic{formel4}c), (\arabic{frequency}c), 
(\arabic{kappa}c) we then obtain 
\begin{equation}
 \frac{d}{dt} P_a(i,t) = P_a(i,t) \sum_j
 M_{ij}^a(t) P_a(j,t)
\label{pers}
\end{equation}
with 
\begin{equation}
\underline{M}^a(t) \equiv \bigg( M_{ij}^a(t) \bigg) 
:= \bigg(\nu_a^3 \kappa_{aa} 
\Big[R_a(i,j;t) - R_a(j,i;t)\Big] \bigg) 
\, . \label{anti}
\end{equation}
Equation (\ref{pers}) has at least the $S_a$ stationary solutions
\begin{equation}
 P_a^s(i) := \delta_{is} = \left\{
\begin{array}{ll}
1 & \mbox{if } i=s \\
0 & \mbox{if } i\ne s
\end{array}\right. \, , \quad s=1,\dots,S_a.
\end{equation}
A linear stability analysis with
\begin{equation}
 \delta P_a^s(i,t) := P_a(i,t) - P_a^s(i)\,, \qquad
 \delta P_a^s(i,t) \ll 1 
\label{linear}
\end{equation}
leads to the equations
\begin{equation}
 \frac{d}{dt} \delta P_a^s(i,t) = \left\{
\begin{array}{ll}
- M_{si}^a \delta P_a^s(i,t) & \mbox{for } i\ne s \\
\sum_j M_{ij}^a \delta P_a^s(j,t) & \mbox{for } i=s
\end{array}\right. 
\end{equation}
(where $\delta P_a^s(i,t) > 0$ for $i \ne s$, $\delta P_a^s(s,t)
= - \sum_{i(\ne s)} \delta P_a^s(i,t) < 0$). Therefore, 
the stationary solution $P_a^s(i)$ is {\em stable} if
\begin{equation}
 M_{si}^a > 0\,, \quad \mbox{i.e.} \quad
 R_a(s,i) > R_a(i,s) \qquad \mbox{for every $i$ with $i\ne s$.}
\label{stable}
\end{equation}
Otherwise it is unstable. Because of $M_{si}^a = -M_{is}^a$ 
(see (\ref{anti})), there is at most one stationary solution $P_a^l(i)$
stable. Attitude $l$ will be the only surviving attitude then. 
If none of the solutions $P_a^s(i)$ is stable, there may exist another
stationary solution $P_a^0(i)$, which is given by
\begin{equation}
 \sum_j M_{ij}^a P_a^0(j) = 0 \quad \mbox{for every $i$.}
\end{equation}
A linear stability analysis with (\ref{linear}) leads to 
\begin{equation}
 \frac{d}{dt} \delta P_a^0(i,t)
= P_a^0(i) \sum_j M_{ij}^a \delta P_a^0(j,t) \, .
 \label{os}
\end{equation}
Since $\underline{M}^a$ is an antisymmetric matrix, i.e.
\begin{equation}
 \bigg( M_{ji}^a \bigg) = - \bigg( M_{ij}^a \bigg)
\label{antisym}
\end{equation}
(see (\ref{anti})), equation (\ref{os}) has only imaginary eigenvalues
({\sc May} (1973)). 
Therefore, an oscillatory behavior of $\delta P_a^0(i,t)$
and $P_a(i,t)$ results. (\ref{antisym}) also implies that
\begin{equation}
 \sum_i \bigg[ P_a(i,t) - P_a^0(i) \ln P_a(i,t) \bigg] = \mbox{const.,} \quad
\mbox{i.e.} \quad \prod_i P_a(i,t)^{P_a^0(i)} = \mbox{const.}
\label{invar}
\end{equation}
(see {\sc May} (1973)). As a consequence of (\ref{invar}) and (\ref{nor}), 
$P_a(i,t)$ moves on a $(S_a-2)$-dimensional
hypersurface (see figures 15, 18 and 19). Especially for $S_a=3$ attitudes,
we find a {\em cyclical} movement. This can be seen in figure 15
($\underline{R}_a = \underline{R}^3$), where simulations for different
initial values are depicted.
\par
Let us examine the case with $\underline{R}_a = \underline{R}^3$ 
in detail. $\underline{M}^a$ is of the form
\begin{equation}
 \bigg( M_{ij}^a \bigg) = \nu_a^3 \kappa_{aa} \left(
\begin{array}{ccc}
0 & -1/2 & 1/2 \\
1/2 & 0 & -1/2 \\
-1/2 & 1/2 & 0
\end{array}\right) \, ,
\end{equation}
then. As a consequence, the stationary solutions $P_a^s(i)$ with $s=1,2,3$
are unstable (see (\ref{stable})). 
$P_a^0(i) \equiv 1/S_a$ is another stationary
solution and has imaginary eigenvalues. Equation (\ref{pers}) takes the form
\begin{equation}
 \frac{d P_a(i,t)}{dt} = \frac{\nu_a^3 \kappa_{aa}}{S_a -1} P_a(i,t) \bigg[
P_a(i-1,t) - P_a(i+1,t) \bigg] \equiv \lambda_i^a(t) P_a(i,t)
\quad \mbox{with} \quad i \equiv i \mbox{ mod } S_a \, .
\label{see}
\end{equation}
$\lambda_i^a(t):= \nu_a^3 \kappa_{aa} 
[P_a(i-1,t) - P_a(i+1,t)]/(S_a - 1)$ is a {\em dynamic
order parameter} and can be interpreted as time dependent {\em growth rate}.
From (\ref{see}) one can see the following:
\begin{itemize}
\item $P_a(i,t)$ increases as long as $P_a(i-1,t) > P_a(i+1,t)$, i.e.
$\lambda_i^a(t)>0$.
\item The growth of $P_a(i,t)$ induces a growth of  $P_a(i+1,t)$ by
$\lambda_{i+1}^a(t) > 0$, and a decrease of $P_a(i-1,t)$ by 
$\lambda_{i-1}^a(t) < 0$.
\item As soon as $P_a(i+1,t)$ exceeds $P_a(i-1,t)$ the decrease of 
$P_a(i,t)$ begins, because of $\lambda_i^a(t) < 0$.
\item These phases are repeated again and again due to $i \equiv
i \mbox{ mod } S_a$ (i.e. a ``cyclical sequence of 
the $S_a$ attitudes'').
\end{itemize}
The above situation can occur, if the readiness to exchange the present
attitude $i$ for the previous attitude $(i-1)$ is less than the readiness to
exchange $i$ for a new attitude $(i+1)$. After a finite number $S_a$ of 
changes one may return to the original attitude, and the process starts from
the beginning. Typical examples are the behavior of consumers 
(concerning fashion) or voters. 
\par
In figures 16 and 17 ($A=1$), the solution of 
equation (\ref{see}) for a varying number $S_1$ of attitudes is depicted.
With growing $S_1$, the temporal development of the fractions
$P_1(i,t)$ becomes more complex. 
For $S_1=5$ attitudes, $P_1(i,t)$ looks already
quite irregular (see fig. 17). However, a {\em phase 
plot}\footnote{Here, the term {\em phase plot} has the meaning 
of a two-dimensional projektion of the trajectory
$(P_1(1,t),\dots,P_A(S_A,t))$. For example, a plot
of the value $P_a(2,t)$ over the value $P_a(1,t)$
with varying time $t$ would be a special case of a phase plot.}
illustrates that $P_1(i,t)$ is still
periodic, since it shows a {\em closed} curve (see fig. 18). A similar
situation can be found for $S_a=6$ attitudes (see fig. 19). For $S_a=7$ 
even the phase plot appears irregular.
\par
Let us return to figure 3, where two subpopulations periodically change
between \mbox{$S_1=4$} resp. $S_2=3$ 
attitudes with different frequencies independently
of each other. The complex ratio between these frequencies can be easily seen
in the corresponding phase plot (see fig. 20: $\underline{\kappa}
= \underline{\kappa}^0_0$). However, if subpopulation $a=2$
is influenced by subpopulation \mbox{$a=1$} (see fig. 4), the phase
plot differs drastically (see fig. 21: $\underline{\kappa} =
\underline{\kappa}^0_1$). Obviously, the oscillations of $P_2(i,t)$
in subpopulation 2 have the same frequency as in subpopulation~1, then.
This remarkable adaption of frequency is called {\em frequency locking}.

\section{Outlook} \label{Out}

We have developed a model for attitude formation by pair interactions, 
which can be written in the form
\begin{equation}
 \frac{d}{dt} P_a(i,t) = \sum_{i'} \bigg[
w^a(i,i';t)P_a(i',t)
- w^a(i',i;t)P_a(i,t) \bigg]
\label{result1}
\end{equation}
with the {\em effective} transition rates
\begin{equation}
 w^a(i',i;t) := R_a(i',i;t)
\sum_b \bigg[ \Big(\nu_{ab}^1(t) - \nu_{ab}^2(t) \Big)P_b(i,t)
+ \nu_{ab}^2(t) + \Big( \nu_{ab}^3(t) - \nu_{ab}^2(t) \Big)
P_b(i',t) \bigg]
\label{result2}
\end{equation}
(see equations (\arabic{formel4}) and (\arabic{frequency})).
This model includes attitude changes by an avoidance behavior, by a readiness
for compromises and by persuasion. It also takes into account different types
of behavior by distinguishing several subpopulations $a$.
\par
The model allows some modifications and
generalizations, which shall be discussed in
forthcoming pubplications:
\begin{itemize}
\item Equations of type (\arabic{formel1})
can produce {\em chaotic}
attitude changes ({\sc Helbing} (1991c)), 
i.e. a temporal behavior that is unpredictable,
since it depends on the initial conditions in a very sensible way
({\sc Schuster} (1984), {\sc Hao} (1984)).
\item Other interesting effects are generated by spatial interactions in 
attitude formation ({\sc Helbing} (1991e)). 
Such spatial interactions may result from migration ({\sc Schweitzer} et. al.
(1991), {\sc Weidlich} \& {\sc Haag} (1988))
or by telecommunication.
\item Attitude formation could be reformulated in a way that allows to 
understand attitude changes as a reaction on a {\em social field}
({\sc Helbing} (1991c), {\sc Lewin} (1951), {\sc Spiegel} (1961)).
\item There are {\em several} plausible specifications of the 
readiness $R_a(i',i;t)$ for attitude changes from $i$ to $i'$: 
\begin{itemize}
\item In the model of this paper, assumptions (\ref{inverse}) and
(\ref{PRO}) imply the relation
\begin{eqnarray}
 \sum_{i'} R_a(i',i;t) &=& \sum_{i'(\ne i)}
 \left( \frac{1}{S_a -1} - R_a(i,i';t) \right) + R_a(i,i;t) \nonumber \\
 &=& 1 - P_{a,i}(t) + P_{a,i}(t) = 1 \, ,
\end{eqnarray}
and $R_a(i',i;t)$
could be interpreted as the {\em probability} 
for an individual of subpopulation~$a$
to change the attitude from
$i$ to $i'$ during a time interval $\tau^a := 1/\nu^a$, where
\begin{equation}
 \nu^a(t) := \sum_b \bigg[ \Big(\nu_{ab}^1(t) - \nu_{ab}^2(t) \Big)P_b(i,t)
+ \nu_{ab}^2(t) + \Big( \nu_{ab}^3(t) - \nu_{ab}^2(t) \Big)
P_b(i',t) \bigg] \, .
\end{equation}
Obviously, $\tau^a$ depends on the
rates of pair interactions of types $k$.
\item Another plausible specification of $R_a(i',i;t)$ 
is given by
\begin{equation}
 R_a(i',i;t) := \frac{\mbox{e}^{U_a(i',t) - U_a(i,t)}}{D_a(i',i;t)}
\end{equation}
with
\begin{equation}
 D_a(i',i;t) = D_a(i,i';t) \quad \mbox{and} \quad D_a(i',i,t) > 0
\end{equation}
(compare to {\sc Weidlich} \& {\sc Haag} (1983)).
That means, the readiness $R_a(i',i;t)$ for an individual of type $a$ 
to change the attitude from $i$
to $i'$ will be the greater, the greater
the difference of the {\em plausibilities (utilities)} 
$U_a$ of attitudes $i'$ and $i$ is, and the smaller the 
{\em incompatibility (distance)} $D_a(i',i;t)$ of 
attitudes $i$ and $i'$ is.
\end{itemize}
The optimal specification of the model (\ref{result1}), (\ref{result2})
can only be found by evaluation of empirical data.
\item Therefore, 
methods have to be developed, which are able to determine the parameters
in the models from (complete or incomplete) empirical data 
({\sc Helbing} (1991b)).
\item A quantitative model for the readiness $R_a(i',i;t)$ to exchange an
attitude $i$ by an attitude $i'$ may be developed, which calculates
the value of $R_a(i',i;t)$ as a function of the {\em relation} of 
attitudes $i$ and $i'$. {\sc Osgood} and 
{\sc Tannenbaum}'s (1955) congruity principle,
{\sc Heider}'s (1946) balance theory and
{\sc Festinger}'s (1957) dissonance theory
have already found valuable results on this problem.
\item The models treated in this paper can be generalized to the situation
illustrated in figure~22, where both pair interactions
and indirect (mean field) interactions play an
important role. In that case, we have to use the transition rates
\begin{equation}
 w(\vec{n}',\vec{n};t) = \left\{
\begin{array}{ll}
n_{\vec{x}}^a \widetilde{w}_a(\vec{x}',
\vec{x};\vec{n};t) & \mbox{if } 
\vec{n}' = \vec{n}_{\vec{x}'}^a{}_{\vec{x}}^a \\
n_{\vec{x}}^a n_{\vec{y}}^b \widetilde{w}_{ab}
(\vec{x}',\vec{y}';\vec{x},\vec{y};\vec{n};t) 
& \mbox{if } \vec{n}' 
=\vec{n}_{\vec{x}'}^{a}{}_{\vec{y}'}^{b}{}_{\vec{x}}^a{}_{\vec{y}}^b \\
0 & \mbox{else}
\end{array} \right.
\end{equation}
when solving the master equation (\ref{master}). The corresponding
approximate equations for the expected fractions $P_a(\vec{x},t)$ are,
if $(N_a-1)/N_a \approx 1$:
\begin{equation}
\frac{d}{dt}P_a(\vec{x},t) = \sum_{\vec{x}'} 
\bigg[ w^a\Big(\vec{x},
\vec{x}';\langle \vec{n} \rangle;t\Big)
P_a(\vec{x}',t) - w^a\Big(\vec{x}',
\vec{x};\langle \vec{n} \rangle;t\Big)P_a(\vec{x},t) \bigg]
\label{gen1}
\end{equation}
with
\begin{equation}
 w^a\Big(\vec{x}',\vec{x};
\langle \vec{n} \rangle;t\Big) :=
w_a\Big(\vec{x}',\vec{x};
\langle \vec{n} \rangle;t\Big) 
+ \sum_{b}
\sum_{\vec{y}} \sum_{\vec{y}'}
w_{ab}\Big(\vec{x}',\vec{y}';
\vec{x},\vec{y};
\langle \vec{n} \rangle;t\Big)
P_b(\vec{y},t)\, ,
\label{gen2}
\end{equation}
\begin{equation}
 w_a := \widetilde{w}_a \, , \qquad 
 w_{ab} := N_b \widetilde{w}_{ab} 
\end{equation}
and
\begin{equation}
 \langle n_{\vec{x}}^a \rangle = N_a P_a(\vec{x},t) \, .
\end{equation}
Equations (\ref{gen1}), (\ref{gen2}) are very general and enable a great
spectrum of quantitative models for social processes: Special cases of
these equations are 
\begin{itemize}
\item the {\em logistic equation} ({\sc Verhulst} (1845), {\sc Pearl} (1924),
{\sc Montroll} \& {\sc Badger} (1974))
and the {\sc Lotka-Volterra} equation ({\sc Lotka} (1920, 1956), 
{\sc Volterra} (1931), {\sc Goel}, {\sc Maitra} \& {\sc
Montroll} (1971), {\sc Hallam} (1986), {\sc Goodwin} (1967))
for population dynamics,
\item the {\sc Fisher-Eigen} equation ({\sc Fisher} (1930), {\sc Eigen} (1971),
{\sc Feistel} \& {\sc Ebe\-ling} (1989)) for the evolution of new 
(behavioral) strategies ({\sc Hel\-bing} (1991)), 
\item the {\em gravity model} ({\sc Ravenstein} (1876), 
{\sc Zipf} (1946)) for migration phenomena, and 
\item the models of {\sc Weidlich} and {\sc Haag}
({\sc Weid\-lich} \& {\sc Haag} (1983, 1988), {\sc Weid\-lich} (1972, 1991)).
\end{itemize}
\end{itemize}

{\bf Acknowledgements:} \\
The author is grateful to Prof. Dr. W. Weidlich and PhD Dr. G. Haag
for valuable discussions, and to the {\sc Volkswagen} foundation
for financial support.

\clearpage
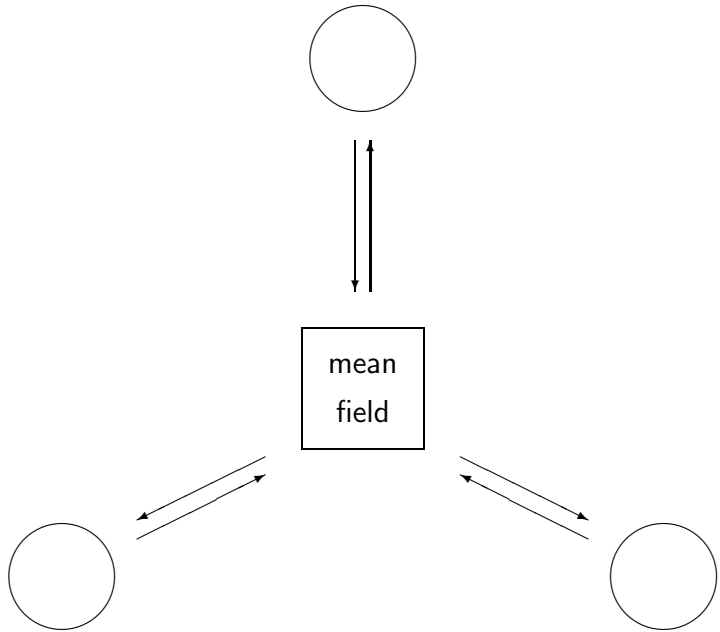
\begin{figure}[htbp]
\unitlength1.0cm
\begin{center}
\begin{picture}(10,7.8)(0,0.8)
\put(1,1){\circle{1.4}}
\put(9,1){\circle{1.4}}
\put(5,7.9){\circle{1.4}}
\put(4.2,2.7){\framebox(1.6,1.6){ }}
\put(4.2,3.3){\makebox(1.6,1){{\sf mean}}}
\put(4.2,2.7){\makebox(1.6,1){{\sf field}}}
%
\put(2,1.5){\vector(2,1){1.7}}
\put(3.7,2.6){\vector(-2,-1){1.7}}
\put(8,1.5){\vector(-2,1){1.7}}
\put(6.3,2.6){\vector(2,-1){1.7}}
\put(4.9,6.8){\vector(0,-1){2}}
\put(5.1,4.8){\vector(0,1){2}}
\end{picture}
\end{center}
\caption[]{Indirect interactions mediated by a mean field for $N=3$ 
individuals.}
\end{figure}
\begin{figure}[htbp]
\unitlength1.0cm
\begin{center}
\begin{picture}(10,7.8)(0,0.8)
\put(1,1){\circle{1.4}}
\put(9,1){\circle{1.4}}
\put(5,7.9){\circle{1.4}}
%
\put(2,1){\vector(1,0){6}}
\put(8,0.8){\vector(-1,0){6}}
\put(1.5,2){\vector(2,3){3.2}}
\put(4.45,6.8){\vector(-2,-3){3.2}}
\put(8.5,2){\vector(-2,3){3.2}}
\put(5.55,6.8){\vector(2,-3){3.2}}
\end{picture}
\end{center}
\caption[]{Direct pair interactions for $N=3$ individuals.}
\end{figure}
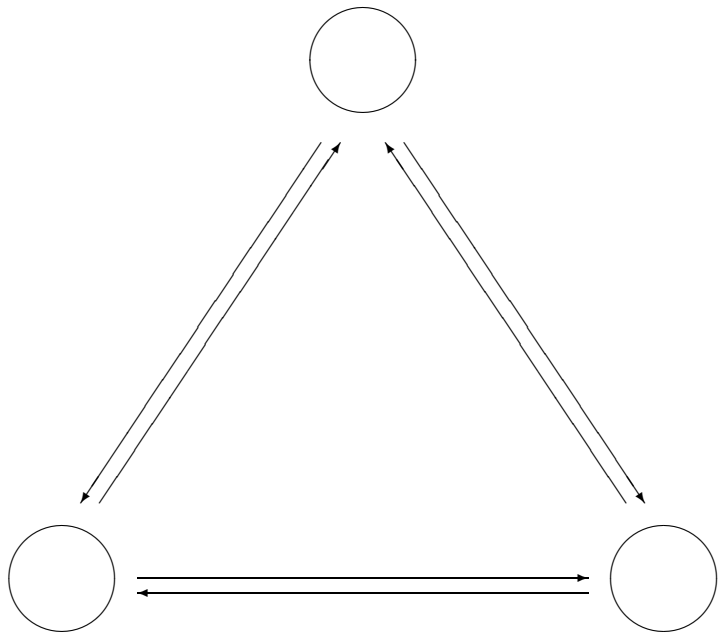
\clearpage
\begin{figure}[htbp]
\caption[]{Two subpopulations changing their attitudes with time $t$
  independently from each other by persuasion. Subpopulation $a=1$ (solid
  lines) changing periodically between 4 attitudes $i$, subpopulation $a=2$
  (broken lines) changing between 3 attitudes $i$ with a different frequency.
{\em (Sorry: Figure is not available as Postscript file. See original publication.)}}
\end{figure}
\begin{figure}[htbp]
\caption[]{Due to asymmetric coupling, changes of attitudes in subpopulation 2
  (broken lines) depend on the attitude distribution in subpopulation 1. In
  contrast to this, subpopulation 1 (solid lines) is not influenced by
  subpopulation 2.
{\em (Sorry: Figure is not available as Postscript file. See original
  publication.)}}
\end{figure}
\begin{figure}[htbp]
\caption[]{Due to mutual coupling, each subpopulation influences attitude
  changes in the other subpopulation. Periodic oscillations of different
  frequencies are destroyed.
{\em (Sorry: Figure is not available as Postscript file. See original
  publication.)}}
\end{figure}
\begin{figure}[htbp]
\caption[]{The equilibrium fractions of attitudes grow with the preferences
  for them. All types of interactions are taken into account.
{\em (Sorry: Figure is not available as Postscript file. See original
  publication.)}}
\end{figure}
\begin{figure}[htbp]
\caption[]{Equal fractions of all attitudes result, if none of the attitudes
  is favoured. All types of interactions are taken into account.
{\em (Sorry: Figure is not available as Postscript file. See original
  publication.)}}
\end{figure}
\begin{figure}[htbp]
\caption[]{Avoidance behavior of two subpopulations prefering the {\em same}
  attitude: Whereas the fractions of the most prefered attitudes are
  decreasing, the factions of the least prefered attitude (dotted line) is 
growing.
{\em (Sorry: Figure is not available as Postscript file. See original
  publication.)}}
\end{figure}
\begin{figure}[htbp]
\caption[]{Avoidance behavior of two subpopulations prefering {\em different}
attitudes: The fraction of the most prefered attitude is growing (solid line).
{\em (Sorry: Figure is not available as Postscript file. See original
  publication.)}}
\end{figure}
\begin{figure}[htbp]
\caption[]{Effects of persuasion and readiness for compromises in the case of
  two subpopulations favouring the {\em same} attitude: Only the attitude with
  the highest preference survives (solid line).
{\em (Sorry: Figure is not available as Postscript file. See original
  publication.)}}
\end{figure}
\begin{figure}[htbp]
\caption[]{Effects of persuasion and readiness for compromises in the case of
  two subpopulataions favouring {\em different} attitudes: A certain fraction
  of individuals decides for a compromise (dotted line).
{\em (Sorry: Figure is not available as Postscript file. See original
  publication.)}}
\end{figure}
\begin{figure}[htbp]
\caption[]{Effects of persuasion in the case of two subpopulations favouring
  the {\em same} attitude: Only the most prefered attitude survives 
(solid line).
{\em (Sorry: Figure is not available as Postscript file. See original
  publication.)}}
\end{figure}
\begin{figure}[htbp]
\caption[]{Effects of persuasion in the case of two subpopulations favouring
  {\em different} attitudes: The attitude favoured by a subpopulation becomes
  the predominating one (solid line), but the attitude which is favoured in
  the other subpopulation also wins a certain fraction (broken line).
{\em (Sorry: Figure is not available as Postscript file. See original
  publication.)}}
\end{figure}
\begin{figure}[htbp]
\caption[]{Periodic oscillations are one possible effect of persuasion. Two
  mutually coupled subpopulations are changing between 3 attitudes each with
  the same frequency and amplitude.
{\em (Sorry: Figure is not available as Postscript file. See original
  publication.)}}
\end{figure}
\clearpage
\begin{figure}[htbp]
\epsfxsize=8.5cm                   
\centerline{\rotate[r]{\hbox{\epsffile[57 40 570 613]
{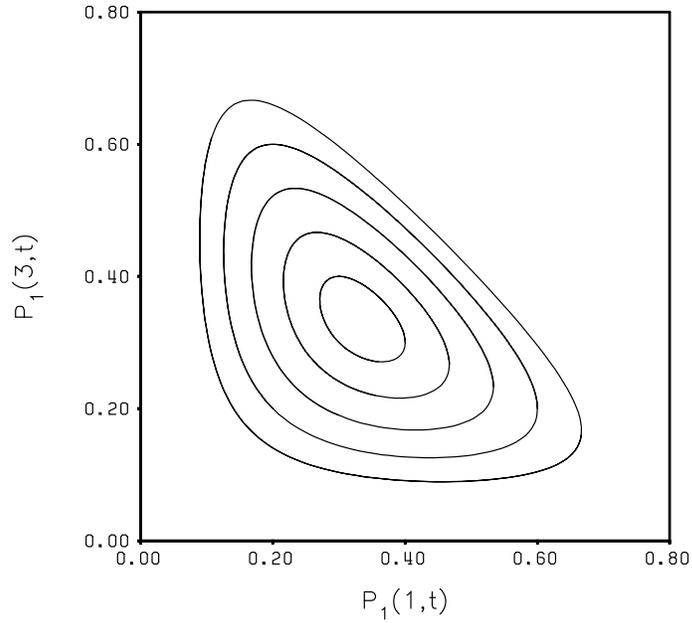}}}}
\caption[]{Phase plots for periodic changes between 3 attitudes in one
  subpopulation: For varying initial conditions different orbits result.}
\end{figure}
\begin{figure}[htbp]
\caption[]{Periodic oscillations in a subpopulation changing between 4
  attitudes by persuasion: Two different amplitudes of the oscillations 
appear. {\em (Sorry: Figure is not available as Postscript file. See original
  publication.)}}
\end{figure}
\clearpage
\begin{figure}[htbp]
\epsfxsize=8.5cm 
\centerline{\rotate[r]{\hbox{\epsffile[57 40 570 756]{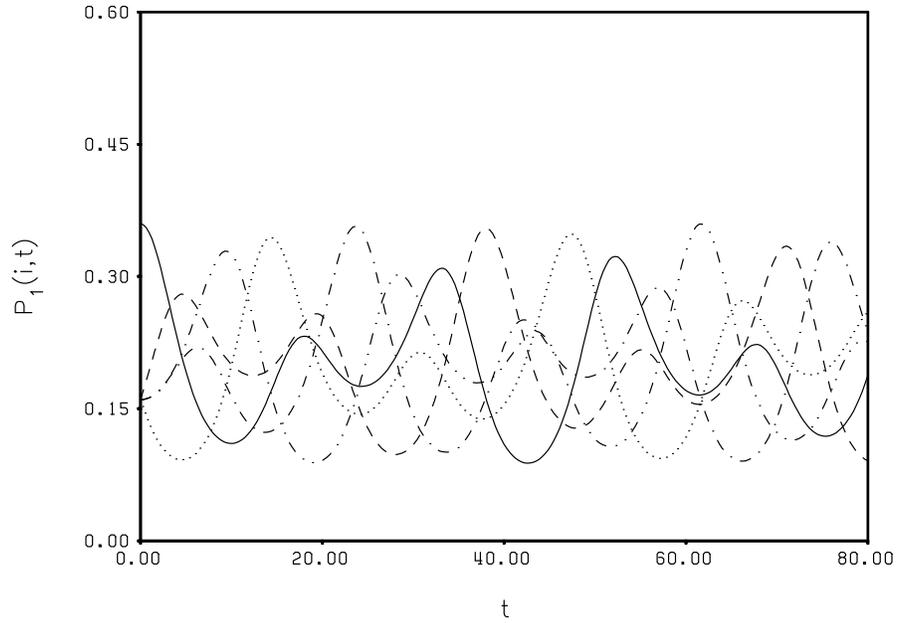}}}}
\caption[]{Oscillatory changes between 5 attitudes show a quite irregular
  temporal development without short-term periodicity.}
\end{figure}
\begin{figure}[htbp]
\epsfxsize=7.8cm                   
\centerline{\rotate[r]{\hbox{\epsffile[57 40 550 613]
{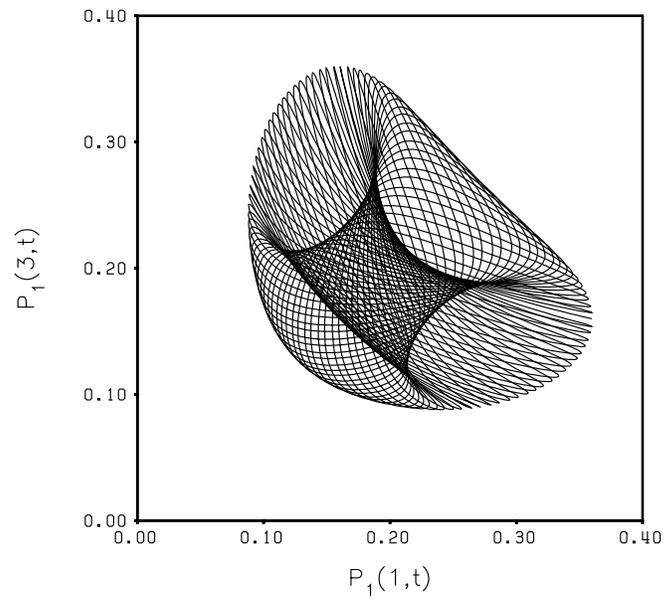}}}}
\caption[]{Phaseplot of oscillatory changes between 5 attitudes shaped like
  the surface of a torus: A long-term periodicity is indicated by the
  closeness of the curve.}
\end{figure}
\clearpage
\begin{figure}[htbp]
\epsfxsize=7.8cm
\centerline{\rotate[r]{\hbox{\epsffile[57 40 550 613]
{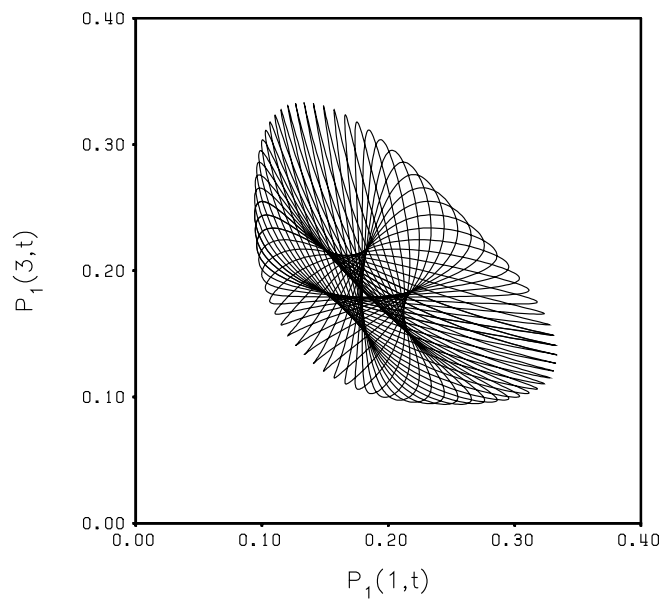}}}}
\caption[]{Phaseplot of oscillatory changes between 6 attitudes indicating a
  longterm periodicity.}
\end{figure}
\begin{figure}[htbp]
\caption[]{Phaseplot for two subpopulations changing between 4 resp. 3
  attitudes independently of each other in a periodic manner and with
  different frequencies.
{\em (Sorry: Figure is not available as Postscript file. See original
  publication.)}}
\end{figure}
\begin{figure}[htbp]
\caption[]{Phaseplot showing an adaptation of frequency by subpopulation 2
  {\em (frequency locking)} as a consequence of the influence of subpopulation
  1 on subpopulation 2.
{\em (Sorry: Figure is not available as Postscript file. See original
  publication.)}}
\end{figure}
\clearpage
\begin{figure}[htbp]
\unitlength1.0cm
\begin{center}
\begin{picture}(10,7.8)(0,0.8)
\put(1,1){\circle{1.4}}
\put(9,1){\circle{1.4}}
\put(5,7.9){\circle{1.4}}
\put(4.2,2.7){\framebox(1.6,1.6){ }}
\put(4.2,3.3){\makebox(1.6,1){{\sf mean}}}
\put(4.2,2.7){\makebox(1.6,1){{\sf field}}}
%
\put(2,1.5){\vector(2,1){1.7}}
\put(3.7,2.6){\vector(-2,-1){1.7}}
\put(8,1.5){\vector(-2,1){1.7}}
\put(6.3,2.6){\vector(2,-1){1.7}}
\put(4.9,6.8){\vector(0,-1){2}}
\put(5.1,4.8){\vector(0,1){2}}
\put(2,1){\vector(1,0){6}}
\put(8,0.8){\vector(-1,0){6}}
\put(1.5,2){\vector(2,3){3.2}}
\put(4.45,6.8){\vector(-2,-3){3.2}}
\put(8.5,2){\vector(-2,3){3.2}}
\put(5.55,6.8){\vector(2,-3){3.2}}
\end{picture}
\end{center}
\caption[]{Indirect interactions mediated by a mean field {\em and} direct
  pair interactions for $N=3$ individuals.}
\end{figure}
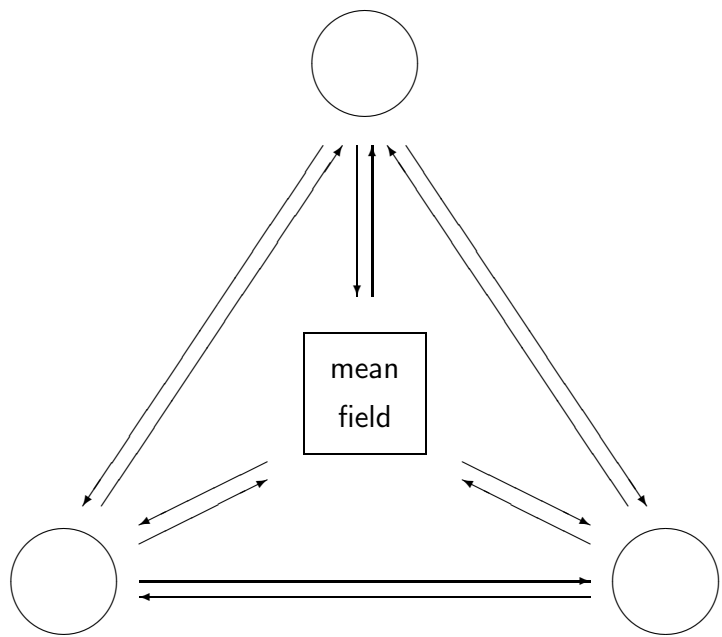
\end{document}